\documentclass[preprint,aps,showpacs,nofootinbib,preprintnumbers,amsmath,amssymb]{revtex4-1}
\usepackage{amssymb}
\usepackage{epsfig}
\usepackage{graphicx}
\usepackage{subfigure}
\usepackage{dcolumn}
\usepackage{bm}
\usepackage[usenames ,dvipsnames]{xcolor}
\usepackage{slashed}

\begin{document}

\title{Revisiting Multi-Component Dark Matter with New AMS-02 Data}

\author{Chao-Qiang~Geng$^{1,2,3}$\footnote{geng@phys.nthu.edu.tw},
Da~Huang$^{2}$\footnote{dahuang@phys.nthu.edu.tw}
and
Chang Lai$^{1}$\footnote{laichang@cqupt.edu.cn}}
  \affiliation{$^{1}$Chongqing University of Posts \& Telecommunications, Chongqing, 400065, China\\
  $^{2}$Department of Physics, National Tsing Hua University, Hsinchu, Taiwan\\
  $^{3}$Physics Division, National Center for Theoretical Sciences, Hsinchu, Taiwan }

\date{\today}
\begin{abstract}
We revisit the multi-component leptonically decaying dark matter (DM) scenario to explain the possible electron/positron excesses 
with the recently updated AMS-02 data. We find that both the single- and two-component DM models can fit the positron fraction 
and  $e^+/e^-$ respective fluxes, in which the two-component ones provide better fits. However, for the single-component models, 
the recent AMS-02 data on the positron fraction limit the DM cutoff to be smaller than 1~TeV, which conflicts with the high-energy 
behavior of the AMS-02 total $e^++e^-$ flux spectrum, while the two-component DM models do not possess such a problem. We also 
discuss the constraints from the Fermi-LAT measurement of the diffuse $\gamma$-ray spectrum. We show that the two-component 
DM models are consistent with the current DM lifetime bounds. In contrast, the best-fit DM lifetimes in the single-component models 
are actually excluded. 
\end{abstract}

\maketitle

\section{Introduction}
\label{s1}
Recently, the AMS-02 collaboration has updated the measurements on the positron fraction~\cite{AMSf} and  electron/positron 
respective fluxes~\cite{AMSep} in the cosmic rays (CRs), which have further confirmed the electron/positron excesses observed 
by the previous experiments, such as AMS~\cite{AMS01,AMS02}, ATIC~\cite{ATIC}, PAMELA~\cite{PAMELA,PAMELA2}, 
and Fermi-LAT~\cite{FermiLAT,FermiLAT1,FermiLATp}.
More interestingly, the new data show some features which have not been observed previously. The most important message is 
that the positron fraction stops increasing with energy~\cite{AMSf}. For the electron/positron fluxes, both spectra become harder at 
$\sim 30$~GeV~\cite{AMSep} so that they cannot be fitted with the usual single power-law functions. Moreover, from 20 to 200~GeV, 
the positron spectral index is larger than the electron one, which indicates that the uprise behavior in the positron fraction originates 
from the hardening of the positron fluxes, a typical hint towards the need for the primary $e^+/e^-$ sources. 
Among the possible primary $e^+/e^-$ origins, pulsars~\cite{pulsar,Lin:2014vja,Boudaud:2014dta} and 
annihilating~\cite{DMindependent,annihilation,annihilationAMS,AnnihilationDecay,Lin:2014vja,Jin:2014ica,Boudaud:2014dta}/decaying~\cite{decay, Ishiwata:2009vx,decayAMS,3bodydecay,3bodydecayAMS,Chen:2009gd,2body,Lin:2014vja,Jin:2014ica,Ko:2014lsa,2bodyAMSa} 
dark matters (DMs) are two popular interpretations extensively studied in the literature. One stringent constraint on the 
DM interpretation is the PAMELA measurement of the anti-proton flux~\cite{PAMELApr}, 
which agrees with the conventional astrophysical prediction very well. 
More recently, AMS-02 Collaboration~\cite{AMSpr} has presented its preliminary measurements of the antiproton to proton ratio as well as 
the latest data on the flux spectra of protons and helium with some new features. 
Nevertheless, in Refs.~\cite{Giesen:2015ufa,Jin:2015sqa} it has been pointed out 
that the new AMS-02 data still accord with the PAMELA ones, which can be explained by the usual secondary antiprotons. 
By fitting with the AMS-02 data, some stronger constraints on the DM models has been given in Refs.~\cite{Giesen:2015ufa,Jin:2015sqa}.
 A simple way to avoid such constraints is to assume that the DMs couple to the Standard Model (SM) only via the lepton sector, which is usually called the leptophilic DM scenario. Note that it was also shown in Ref.~\cite{astrophysical} that the anomalous behavior of positron flux could be possibly explained within the conventional astrophysical framework by introducing some unconventional positron secondary production mechanisms and non-standard propagation models\footnote{We mention that the results in the first paper of Ref.~\cite{astrophysical} was recently questioned by Ref.~\cite{Dado:2015lta}.}.  

Before the recent release of the AMS-02 data, the AMS-02 positron fraction published last year~\cite{AMS02} and the Fermi-LAT total $e^+ + e^-$ flux~\cite{FermiLAT} represented two of the most precise measurements of the CRs. However, the simplest scenario in which a single DM component annihilating or decaying into lepton pairs cannot fit these two datasets simultaneously~\cite{2bodyAMSa}. In Refs.~\cite{2DM_1,2DM_2}, we have proposed a multi-component DM scenario~\cite{2DM_others} in order to overcome this difficulty. In particular, two DM components with the heavy DM decaying solely to $\mu^+\mu^-$ and the light one predominantly to  $\tau^+\tau^-$ with the energy cutoff at $E_{cL}=100$ GeV could already provide a good fit to the combined dataset of the AMS-02 positron fraction and  Fermi-LAT total $e^++e^-$ flux. As a result, this two-component DM model can explain the apparent substructure at around 100 GeV in  both spectra as the light DM drops at that energy. Another advantage of this multi-component scenario is that it gives us a mechanism to evade the strong DM lifetime bound from the diffuse $\gamma$-ray spectrum measured recently by Fermi-LAT~\cite{FermiLAT_Gamma}, which has already greatly constrained the simple two-body leptonically decaying DM models. 
 We have also checked that the addition of the HESS total $e^+ + e^-$ data~\cite{HESS} in the fitting would not change this general conclusion.

In the light of the updated data from AMS-02, it is useful and necessary to revisit the single- and two-component DM models. 
More remarkably, the new data from AMS-02 still show the substructure around 100~GeV, which strengthens our confidence of the investigation of the multi-component DM scenario. In this work, we shall only use the latest AMS-02 measurements of the positron fraction and
 fluxes of $e^-$ and $e^+$ in our fitting procedure. In this way, we can avoid many systematic uncertainties involved in the AMS-02/Fermi-LAT combined dataset~\cite{Lin:2014vja}, due to the differences in the experiment designs, detector responses and data-taking periods in the solar cycle. Thus, we expect that the final fitting result should be more consistent, which is another motivation for the present work.

The paper is organized as follows. In Sec.~\ref{s2}, we  briefly introduce our multi-component decaying DM models and the propagation physics 
of CRs in the Galaxy. The fitting results about the single- and two-component DM models are presented in Sec.~\ref{s3}. In Sec.~\ref{s4}, we discuss the Fermi-LAT diffuse $\gamma$-ray constraints on these
models. Finally, we  give a short summary in Sec.~\ref{s5}.

\section{Signals and Backgrounds}
\label{s2}
In our multi-component DM framework, the total electron flux is composed of primary, secondary and DM-decay-induced electrons, 
while only secondary positrons and the ones from the DM decays contribute to the total positron flux, which can be written as follows:
\begin{eqnarray}
\Phi^{(\rm{tot})}_{e} &=& \kappa_1 \Phi^{(\rm{primary})}_e + \kappa_2 \Phi^{(\rm{secondary})}_e+\Phi^{\rm{DM}}_e , \nonumber\\
\Phi^{(\rm{tot})}_p &=& \kappa_2 \Phi^{(\rm{secondary})}_p + \Phi^{\rm{DM}}_p.
\end{eqnarray}
The primary electrons are widely believed to be generated from the supernova remnants distributed in our Galaxy~\cite{SNR}, and the
 injection spectrum is usually assumed to be a broken power-law function with respect to the rigidity $\rho$. {Here, we choose the reference electron primary injection spectrum to be the three-piece broken power law:  $q^{e}(\rho) \propto \left(\rho/\rho_{e1,2} \right)^{-\gamma_{e1,2,3}}$, where $\rho_{e1,2}$ refer to the two reference rigidities and $\gamma_{e1,2,3}$ the three spectral indices with 
the relevant parameters shown in Table~\ref{parameters}.} Note that we insert a parameter $\kappa_1$ to account for the normalization uncertainty in the primary electrons. Secondary electron/positron fluxes $\Phi^{(\rm{secondary})}_{e,p}$ are the final products of the collisions of the charged particles in the CRs, such as protons and other nuclei, with the interstellar medium (ISM) in the Galaxy. {In the present work, we follow the diffusion-reacceleration CR propagation model, in which the spatial diffusion coefficient is parameterized as a power law $D_{xx}=\beta D_0 (\rho/rho_r)^\delta$ with $\rho_r$ the reference rigidity, $\beta=v/c$ the velocity and $\delta$ the power spectral index. The reacceleration process is described by the diffusion coefficient in momentum space $D_{pp}=4 v_A^2 p^2/(3D_{xx}\delta(4-\delta^2)(4-\delta))$. The primary CR proton spectrum is also assumed to follow a broken power-law function: $q^n(\rho) \propto (\rho/rho_n)^{\gamma_{n1,2}}$. To concretely compute the CR spectra,} we use the GALPROP code~\cite{GALPROP} to simulate the productions and propagations of these background electrons and positrons with the fixed diffusion coefficients and primary proton parameters shown in Table~\ref{parameters}.
\begin{table}
\caption{Parameters for the diffuse propagation, primary electrons, and primary protons. }
\begin{tabular}[t]{|cccc|ccccc|ccc|}
\hline
\multicolumn{4}{|c|}{diffuse coefficients}&\multicolumn{5}{c|}{primary electrons}
&\multicolumn{3}{c|}{primary protons} \\
\hline
$D_0(\mathrm{cm}^2\mathrm{s}^{-1})$ & $\rho_r(\rm{GV})$  &$\delta$ &$v_A({\rm km\, s}^{-1})$&$\rho_{e1}(\rm{GV})$& $\rho_{e2}(\rm{GV})$ & $\gamma_{e1} $&$\gamma_{e2}$ & $\gamma_{e3}$
&$\rho_{n}(\rm{GV})$&$\gamma_{n1} $&$\gamma_{n2}$\\
\hline
$5.3\times10^{28}$& $4.0$  & 0.33 & $33.5$
&$4.0$ & 67.6 & 1.46 & 2,72 & 2.6 & $11.5$ &1.88 &2.39 \\
\hline
\end{tabular}\label{parameters}
\end{table}
For other details of the calculation, especially the choice of the astrophysical parameters, we refer to our earlier work in Ref.~\cite{2DM_1}. 
However, the calculation of secondary $e^-/e^+$ fluxes involves the uncertainties from, for instance, nuclei collision cross sections, 
form factors of heavy nuclei, and  propagation coefficients, which are partially taken into account with the parameter $\kappa_2$ to rescale the calculated secondary fluxes~\cite{Lin:2014vja}. The parameters $\kappa_{1,2}$ will be determined with other model parameters in the following fitting procedure.

As for the DM signal $\Phi^{\rm DM}_{e,p}$, we assume that the whole DM density in the Galaxy and  Universe is carried out by a single- or multiple-component DM particles $\chi_i$, whose decays can explain the positron/electron anomalies. The dominant decay channels for all DM components are taken to be
\begin{equation}\label{DecayChannel}
\chi_i \to l^{\pm} Y^{\mp},
\end{equation}
where $l=e,\mu$ and $\tau$, and $Y$ is another new charged particle whose further decay is irrelevant to our following discussion. This decay mode can be easily embedded into a full-fledged particle physics model. For example, it is possible that $Y^\pm$ can decay into its neutral partner $Y^0$ plus charged leptons. {If mass difference between $Y^\pm$ and $Y^0$ is less than 100~MeV, the corresponding $e^\pm$ signal is too soft to affect the high-energy $e^\pm$ spectra we are interested in, and the energy released to the early Universe is so limited that its effect on the CMB power spectrum is also suppressed~\cite{CMBpower}.} Such decay channel naturally realizes the leptophilic scenario so that it can satisfy the PAMELA constraint on the antiproton~\cite{PAMELApr}.
The $e^+/e^-$ source terms $Q_{e,p}^{\rm DM}$ induced by the DM decays can be parametrized as:
\begin{eqnarray}\label{source}
Q^{\rm DM}_{e,p} ({\bf x}, p) = \sum_i \frac{\rho_i ({\bf x})}{\tau_i M_i} \left(\frac{dN_{e,p}}{dE}\right),
\end{eqnarray}
where $M_i$, $\tau_i$ and $\rho_i(\bf x)$ are the mass, lifetime and energy density distribution for the $i$-th DM component, respectively. For simplicity, we assume that each DM component carries the same fraction of the entire energy density, so that {$\rho_i({\bf x}) = \rho({\bf x})/N$}, where $\rho(\bf x)$ is the DM density distribution in the Galaxy as {the widely-used NFW profile~\cite{NFW}}. Here, $(dN_{e,p}/dE)_i$ is the differential $e^-/e^+$ multiplicity for each annihilation,  given by the mixture of the three leptonic channels:
\begin{eqnarray}\label{Norm_Spectrum}
\Big(\frac{dN_{e,p}}{dE}\Big)_i=\frac{1}{2}\Big[\epsilon^e_i\Big(\frac{dN^e} {dE}\Big)_i+\epsilon^\mu_i\Big(\frac{dN^\mu}{dE}\Big)_i+\epsilon^\tau_i\Big(\frac{dN^\tau}{dE}\Big)_i\Big]\;,
\end{eqnarray}
where $\epsilon^{e,\mu,\tau}_i$ denote the corresponding branching ratios satisfying the normalization condition $\epsilon^e_i + \epsilon^\mu_i + \epsilon^\tau_i =1$ and the factor $1/2$ takes into account that $e^+$ and $e^-$ are generated in two separated channels. Since the decay channels shown in Eq.~(\ref{DecayChannel}) are all two-body processes, we can easily determine the normalized injection spectrum for each decay process only by the kinematics. Concretely, the injection spectra for $e$- and $\mu$-channels can be calculated analytically,
\begin{eqnarray}
\Big( \frac{dN^e}{dE} \Big)_i &=& \frac{1}{E_{ci}}\delta(1-x),\\
\Big( \frac{dN^\mu}{dE} \Big)_i &=& \frac{1}{E_{ci}}[3(1-x^2)-\frac{4}{3}(1-x)]\theta(1-x),
\end{eqnarray}
with $x=E/E_{ci}$, while the $\tau$-channel spectrum is simulated with PYTHIA~\cite{PYTHIA} due to the complicated $\tau$ hadronic decays. $E_{ci}$ is the energy cutoff of $e^\pm$ for each DM component, and can be determined as follows:
\begin{eqnarray}\label{cutoff}
E_{ci} = \frac{M_i^2-M_Y^2}{2M_i}\, .
\end{eqnarray}
The propagation of electrons and positrons between the DM $e^-/e^+$ sources and the Earth is very complicated~\cite{DiffuseEquation}, which involves the deflection of $e^-/e^+$ in the galactic magnetic fields and energy loss via the inverse Compton (IC) scattering, bremsstrahlung and synchrotron radiation. In this work, such a sophisticated propagation is consistently solved by the GALPROP codes with the same set of diffusion coefficients as 
the background fluxes shown in Table~\ref{parameters}. Finally, it is generally believed that the solar modulation affects the $e^-/e^+$ flux spectra greatly, especially at energies below and around 10 GeV. {But our focus here is the high energy range which is known to be less impacted by this solar modulation. Therefore, we follow the simple force-field approximation~\cite{SolarModulation} with the Fisk potential $\phi_F=0.55$~GV. Note that the choice of this fixes value of Fisk potential is just for illustration, rather than guaranteeing the spectra at energies smaller than 10 GeV to be followed by the our fit. }

It is well-known that the computation of the spectra of various CR particles always suffer from many astrophysical uncertainties, such as the specific values of diffusion coefficients and the choice of DM halo profiles. However, the purpose of the present paper is to investigate the viability of the multi-component DM scenario in light of the new AMS-02 data. Thus, we ignore such complicated issue involving astrophysical uncertainties, and only fix the astrophysical parameters to the specific values in Table~\ref{parameters}. {It is also expected that the use of other DM halo profiles should not modify our general results much, since only $e^\pm$ generated within the local region of about $1$~kpc around the Sun can contribute to the signal. We have checked this statement with the isothermal profile~\cite{isothermal}.}

\section{Fitting Results}
\label{s3}
The datasets used in our study include the latest AMS-02 measurements of the positron fraction~\cite{AMSf} 
and electron and positron respective fluxes~\cite{AMSep}. These three groups of the data may correlate to each other, 
as the positron fraction can be calculated from positron and electron fluxes. Nevertheless, since they have different systematic uncertainties,  
we adopt all of them simultaneously in our fitting procedure. Furthermore, we restrict to the data with the energy above 10 GeV in order to 
reduce the effects of the solar modulation. Thus, we have totally 140 data points. For the fitting procedure, we use the simple $\chi^2$-minimization method to obtain the best-fit point and assess the goodness of the fit. In the following two subsections, we present the fitting results 
for the single- and two-component decaying DM models, which are the simplest ones in the general multi-component DM scenario. 
After fixing the best-fit model parameters, we can predict the total $e^+ + e^-$ flux spectrum and compare it
with the latest measurement by AMS-02~\cite{AMSt}.

\subsection{Results for Single-Component Dark Matter Models}\label{DM1}
In this section, we focus on the simplest case with a single DM component. In order to obtain the meaningful physical results, we fix the DM mass to be $M=3030$~GeV. Therefore, we have totally 6 parameters: the primary and secondary normalization factors $\kappa_1$ and $\kappa_2$, energy cutoff $E_c$, DM lifetime $\tau$ and two independent decay branching ratios $\epsilon^e$ and $\epsilon^\tau$, together with the constraint $\epsilon^e + \epsilon^\tau \leqslant 1$. In order to simplify the fitting procedure, we fix the cutoff $E_c$ to be 600, 800, 1000 and 1500~GeV, respectively, and fit other five parameters for each $E_c$.
{
\begin{table}[ht]
\caption{Parameters leading to the minimal values of $\chi^2$ with the cutoff of the single DM being 600, 800, 1000 and 1500~GeV, respectively.}
{\begin{tabular}{@{}ccccccccc@{}} \toprule
$E_{c}({\rm GeV})$& $\kappa_1$ & $\kappa_2$ & $\epsilon^{e}$& $\epsilon^{\mu}$  & $\epsilon^{\tau}$
 & $\tau(10^{26}{\rm s})$ & $\chi^2_{\rm min}$ & $\chi^2_{\rm min}/{\rm d.o.f.}$
\\
\colrule
600 & 0.94 & 1.60 & 0.07  & 0 & 0.93 & 0.43 & 115 & 0.85 \\
800 & 0.94 & 1.62  & 0.02  &  0 & 0.98  & 0.47 & 128 & 0.95 \\
1000 & 0.94 & 1.65  & 0 &  0 & 1 & 0.51 & 145 & 1.08 \\
1500 & 0.94 & 1.65  & 0  & 0.15 & 0.85 & 0.54 & 215 & 1.60\\
\botrule
\end{tabular}\label{tab_1DM}}
\end{table}
}
\begin{figure}[ht]
\centering
\includegraphics[width=0.33\textwidth, angle =-90]{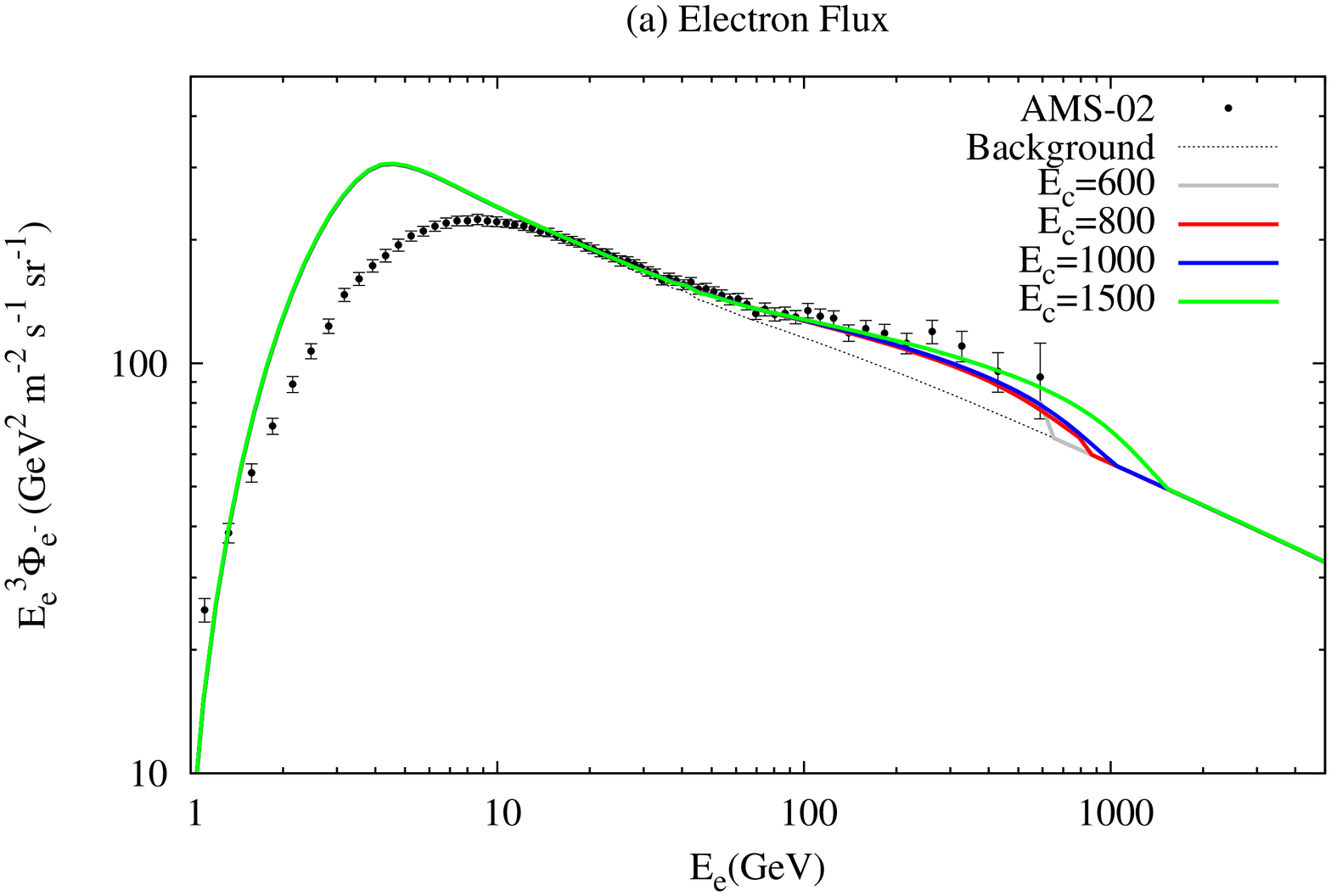}
\includegraphics[width=0.33\textwidth, angle =-90]{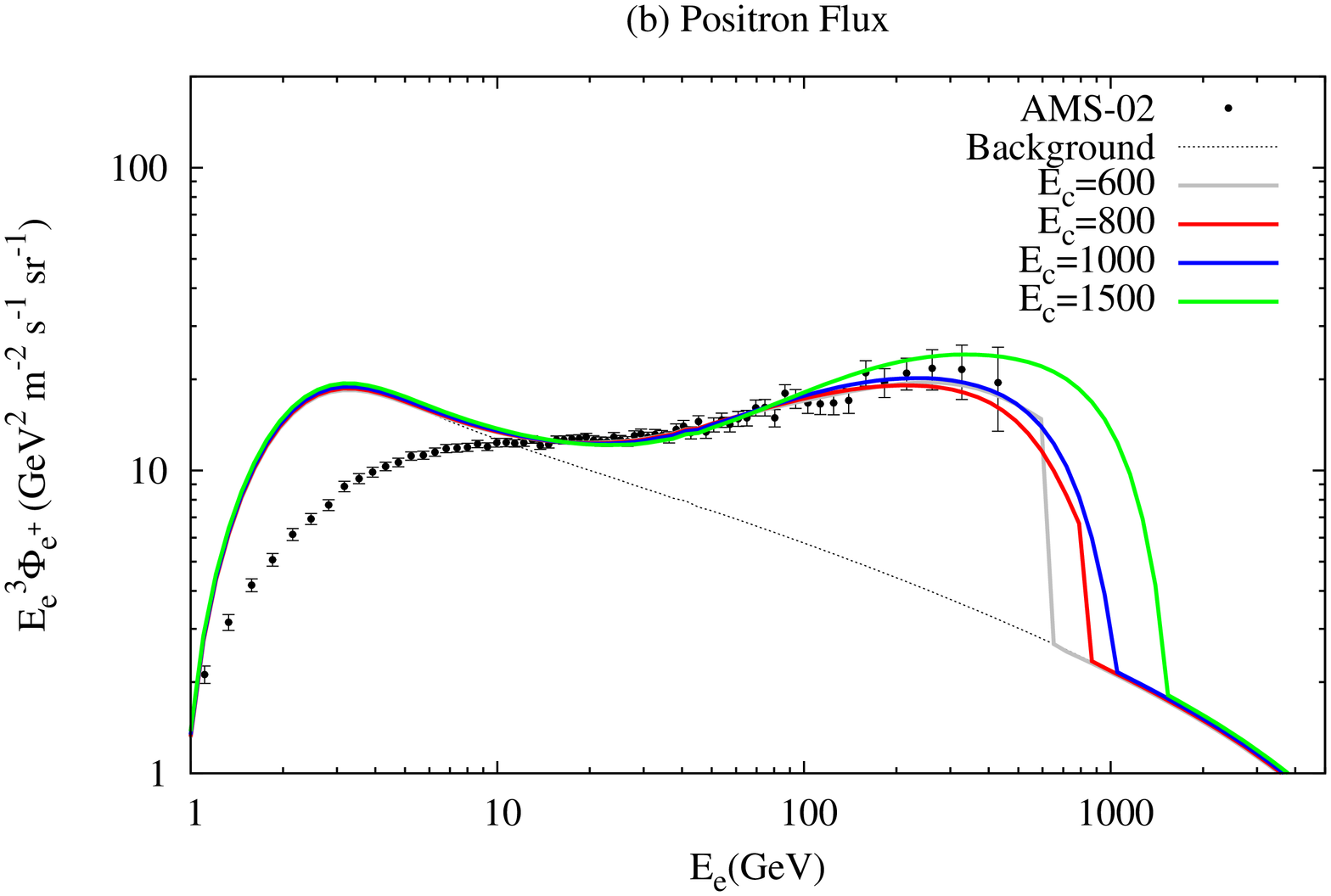}
\includegraphics[width=0.33\textwidth, angle =-90]{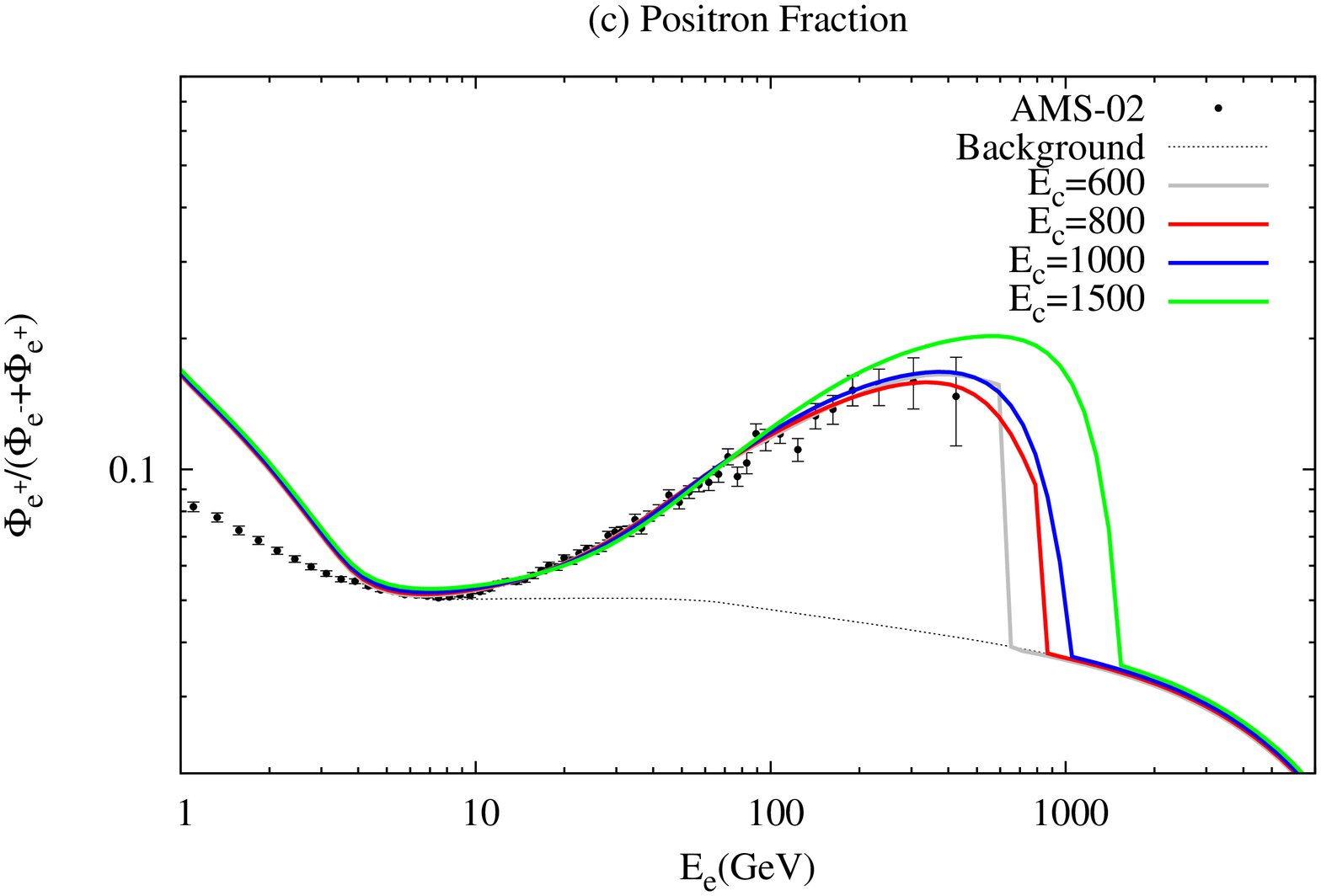}
\includegraphics[width=0.33\textwidth, angle =-90]{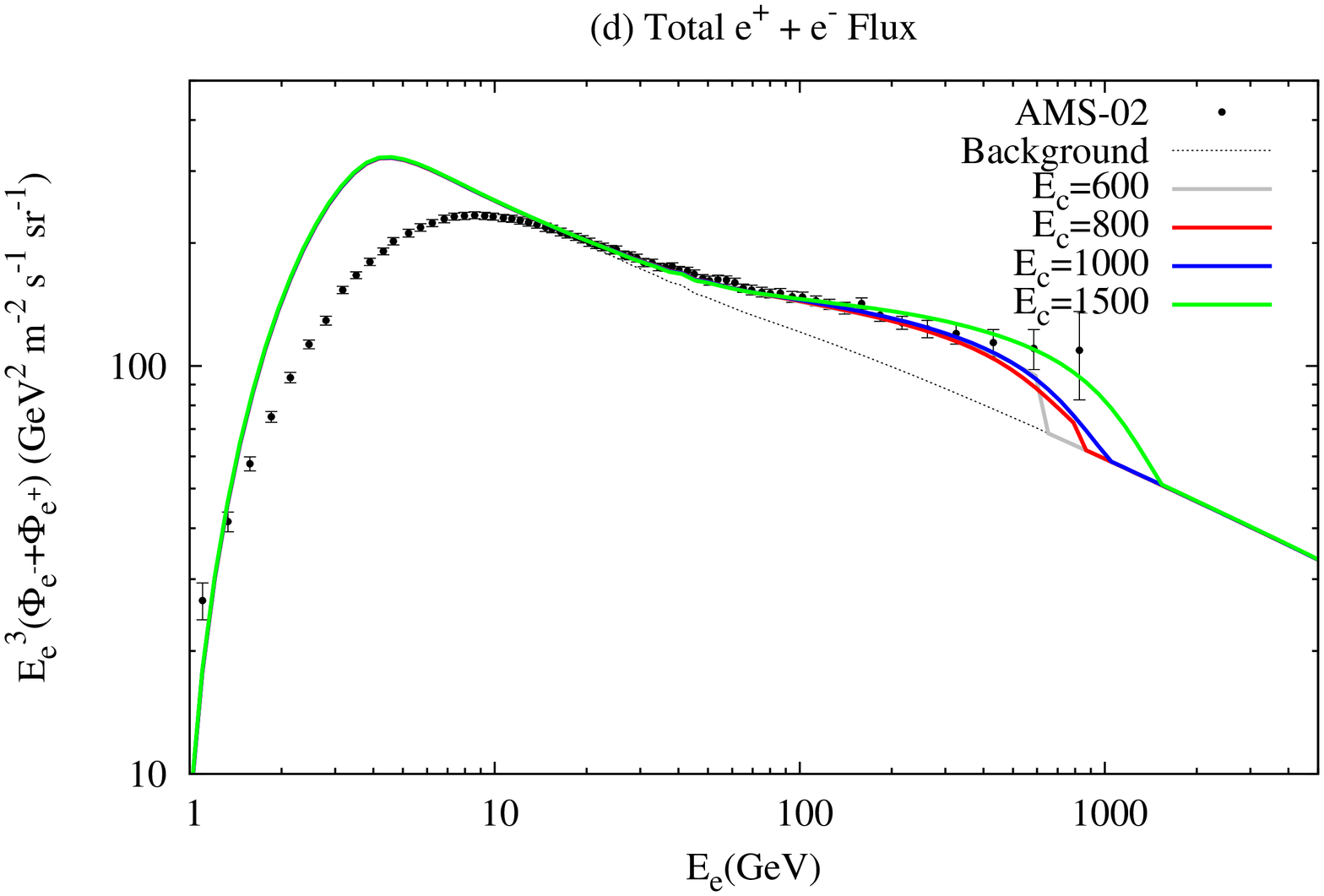}
\caption{(a) Electron flux, (b) positron flux, (c) positron fraction, and (d) total $e^+ + e^-$ flux from the single-component DM contributions with the best-fitting parameters given in Table~\ref{tab_1DM} for $E_{cH}=$600, 800, 1000 and 1500 GeV, respectively.
}\label{Fig_1DM}
\end{figure}

The best-fit results are summarized in Table~\ref{tab_1DM} and Fig.~\ref{Fig_1DM} for different energy cutoffs. From Table~\ref{tab_1DM}, we find that for the first three cases with the energy cutoff smaller than 1~TeV, the single-component DM model can already give good fits to the AMS-02 measurements of the positron fraction and  $e^+/e^-$ respective fluxes, while the last benchmark with $E_c = 1.5$~TeV is not very reasonable due to the too large value of $\chi^2_{\rm min}/{\rm d.o.f.}$
 Note that in Fig.~\ref{Fig_1DM}(d), we show the predictions of the total $e^+ + e^-$ flux with the best-fit parameters. 
 By comparing these predictions with the latest AMS-02 data on the total $e^+ + e^-$ flux, we find that the 
  $e^+ + e^-$ spectrum for $E_c \geqslant 1~$TeV either stops too early or decays too fast, so that it cannot follow the measured high energy behavior, 
 especially for the data with energies larger than 400~GeV. In contrast, the case with $E_c = 1.5$~TeV can give a good description at the 
 high energy, though it proves a bad fit for the other three datasets. From this point of view, the single-component DM models encounter some problems: the AMS-02 data for the positron fraction and $e^+/e^-$ fluxes seem to favor a DM with its cutoff smaller than 1~TeV, but such  a DM makes the total $e^+ + e^-$ flux at the high energy region difficult to be explained.

\subsection{Results for Two-Component Dark Matter Models}\label{DM2}
We now turn to the two-component DM case, in which we use ${\rm DM}_{L(H)}$ to represent the light (heavy) DM. 
Note that we want to explain the substructure around 100~GeV in terms of the light DM stopping to decay at the energy, 
resulting in that the cutoff $E_{cL}$ of ${\rm DM}_{L}$ is fixed to be 100 GeV. However, the cutoff $E_{cH}$ of the heavy DM is free, 
which is taken to be 600, 800, 1200, 1500~GeV in our numerical investigations, respectively. Here, we choose the mass of the heavy particle 
$Y$ to be 300~GeV for simplicity, {so that the two DM masses can be determined via Eq.~(\ref{cutoff}) to be $M_L=416~$GeV and $M_H=1271$, 1654, 2437, and 3030~GeV, respectively.}

\begin{figure}
\centering
\includegraphics[width=0.33\textwidth, angle =-90]{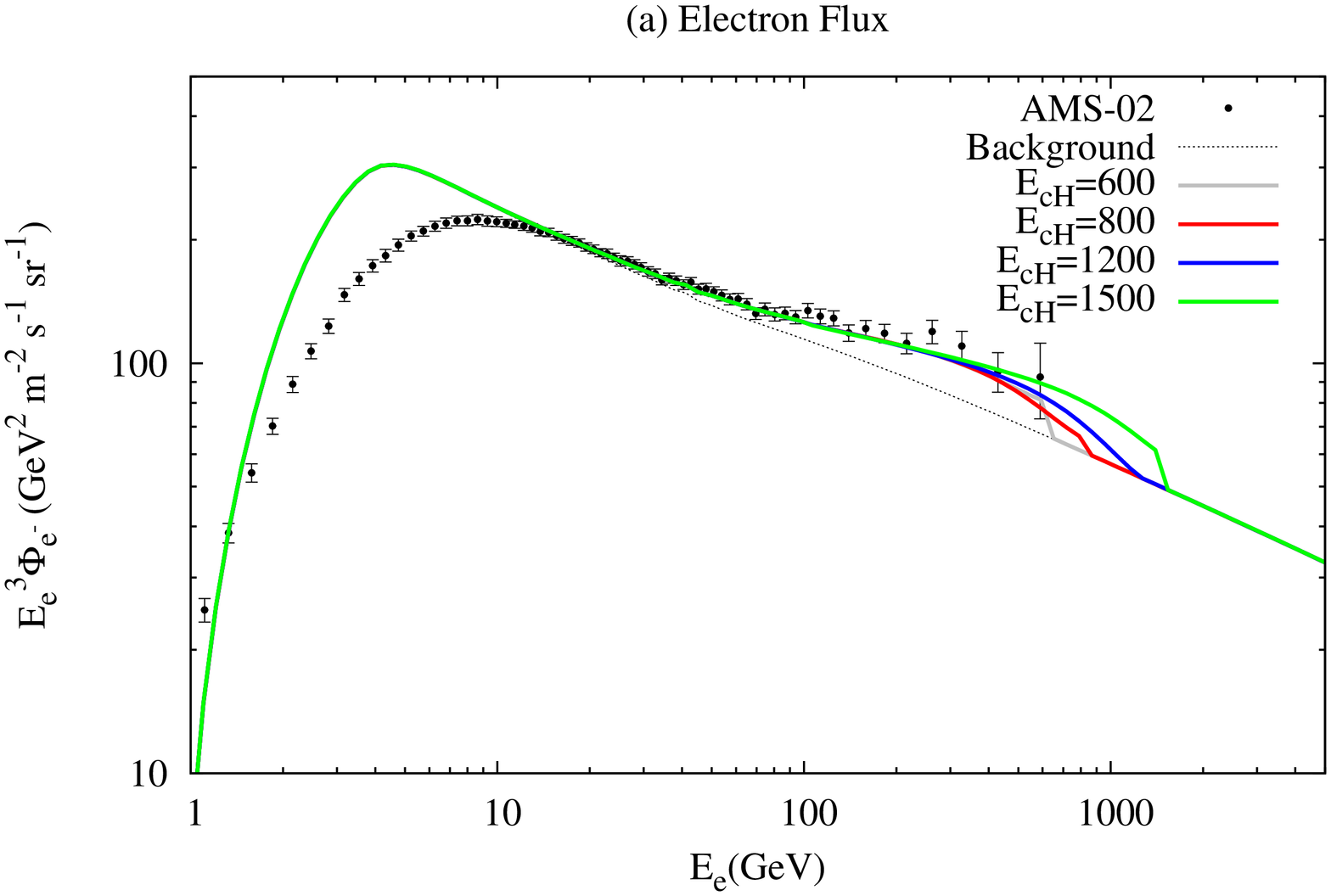}
\includegraphics[width=0.33\textwidth, angle =-90]{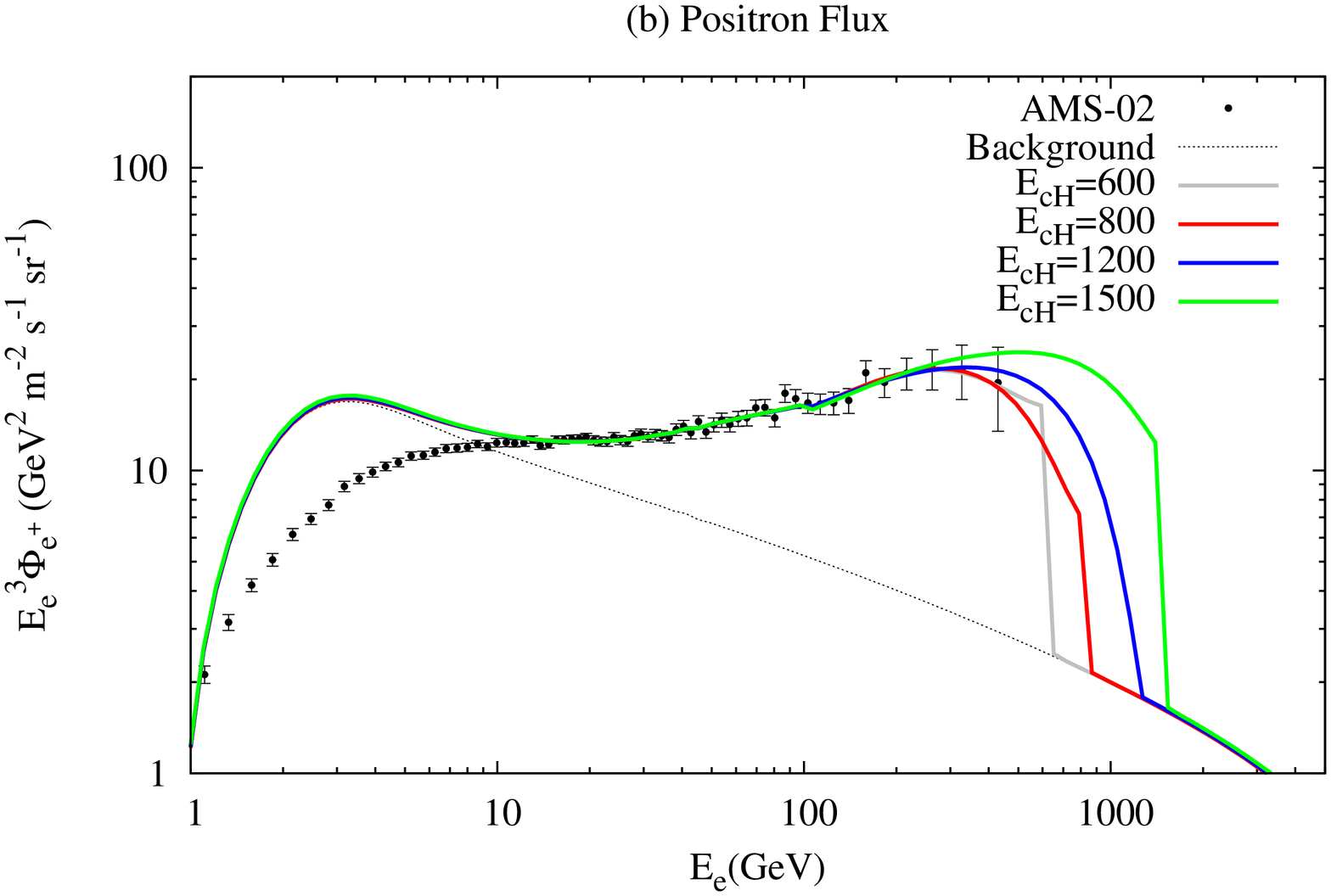}
\includegraphics[width=0.33\textwidth, angle =-90]{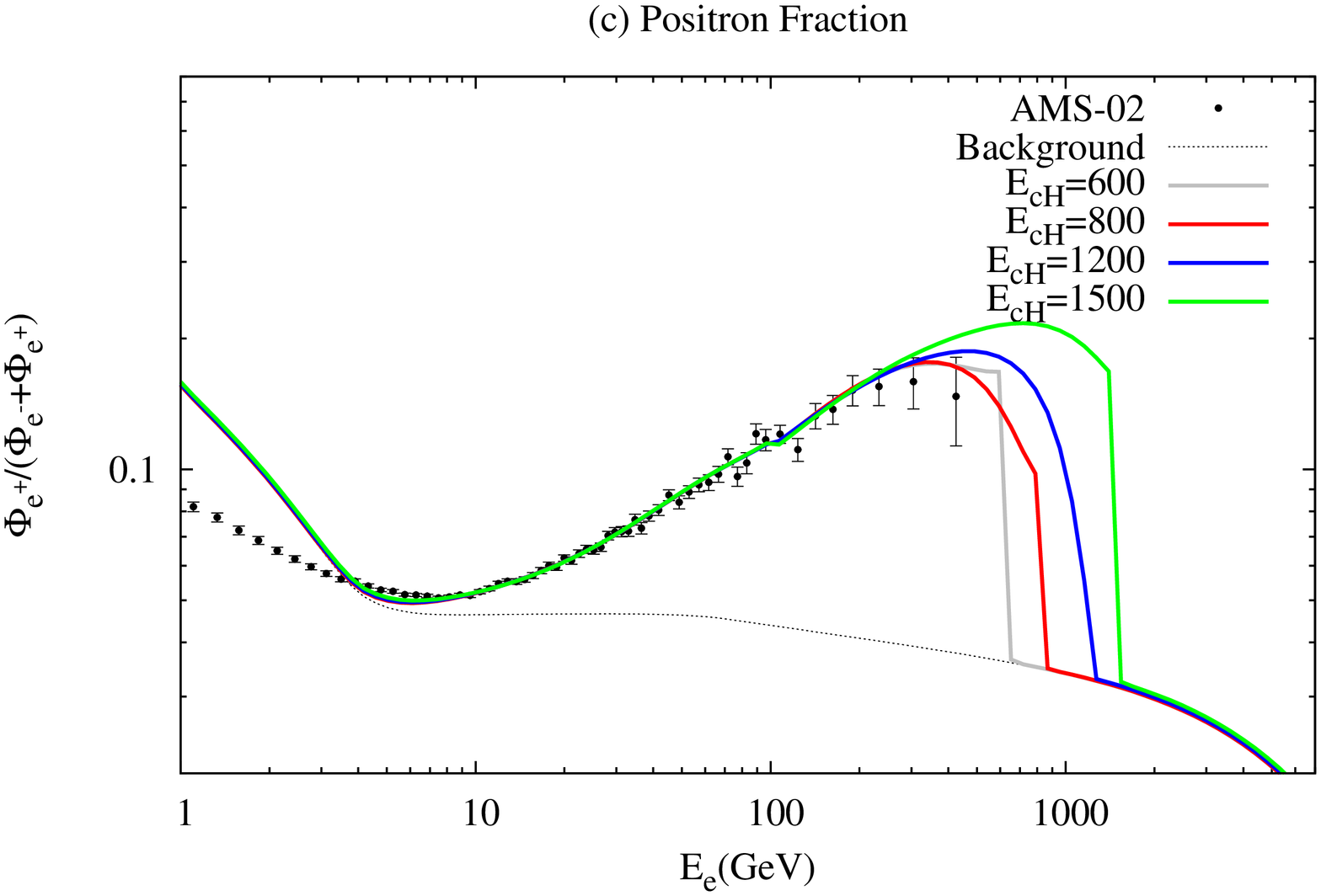}
\includegraphics[width=0.33\textwidth, angle =-90]{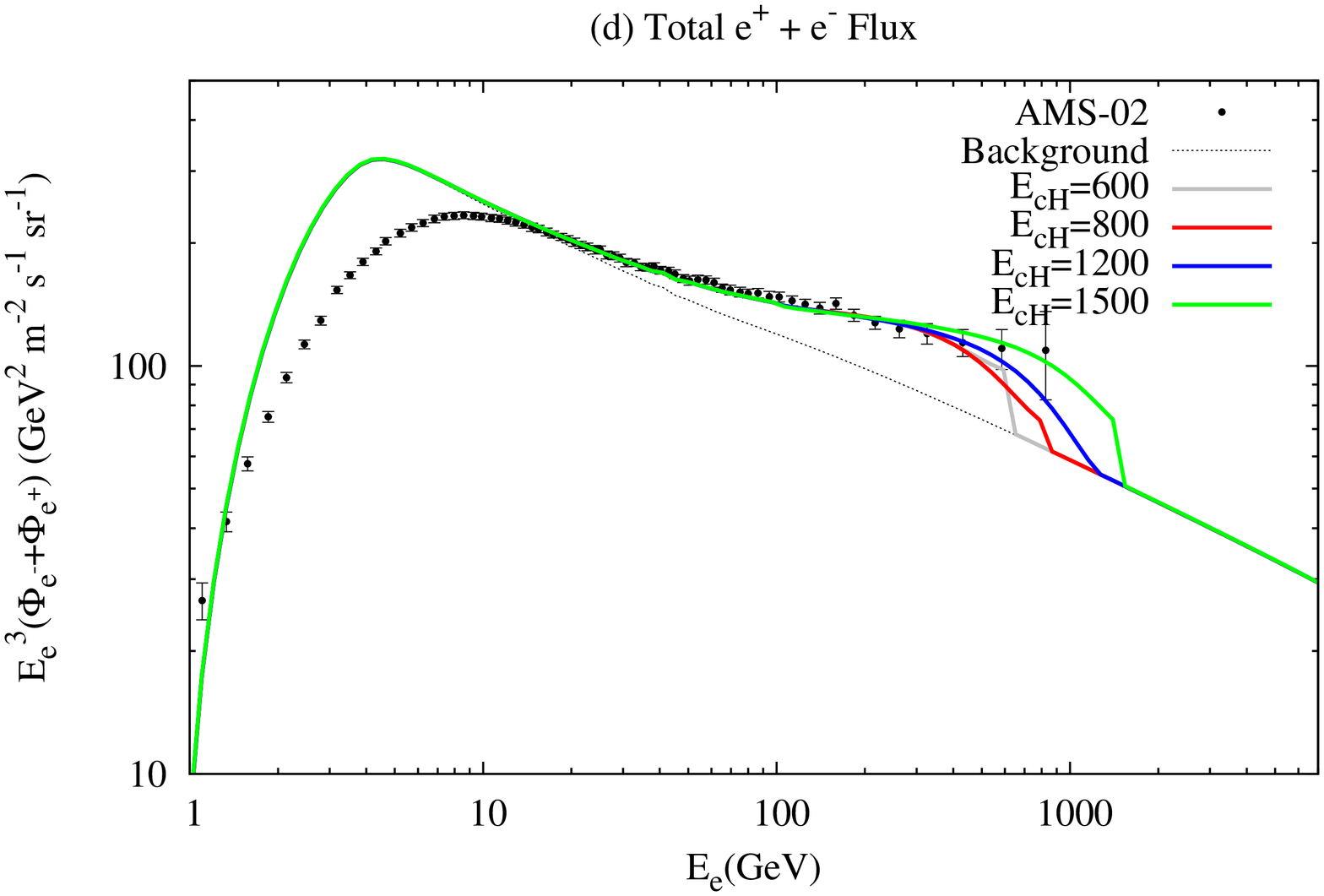}
\caption{(a) Electron flux, (b) positron flux, (c) positron fraction, and (d) total $e^+ + e^-$ flux from the two-component DM contributions with the best-fitting parameters given in Table~\ref{tab_2DM} for $E_{cH}=$600, 800, 1200 and 1500 GeV, respectively.
}\label{Fig_2DM}
\end{figure}

The fitting results are presented in Table~\ref{tab_2DM}, and the predictions with the best-fit parameters are shown 
in Fig.~\ref{Fig_2DM}. Generally speaking, all of the four two-component DM models can fit to the AMS-02 data pretty well 
as $\chi^2_{\rm min}/{\rm d.o.f.} <1$, which are much better than any single-component DM model considered in the previous subsection. 
The flavor structures are almost the same in these cases, in which the heavy DM decays primarily through the $\mu$-channel, 
while the light one favors the $\tau$-channel. The hardening feature observed in the $e^+/e^-$ flux spectra around 30~GeV 
is explained by the transition from the background-dominated region to the DM-dominated one. Even better, the positron fraction 
spectrum with $E_{cH}=800~$GeV shows the start of the decreasing behavior with the maximum at around 300~GeV, which coincides with 
the striking claim in Ref.~\cite{AMSf}. Unfortunately, the predicted total $e^+ + e^-$ flux spectrum for this heavy DM cutoff goes back to the background level too early as compared with the most recent AMS-02 data, giving a bad description to the last two points. In contrast, the spectra with $E_{cH}= 1200$ and 1500~GeV can reduce or solve this problem by extending the DM $e^+ +e^-$ flux to high energies. However, in the latter two cases, the increasing behaviors in the positron fraction also continue to high energies, already exceeding 500~GeV, which disagrees with the conclusion in Ref.~\cite{AMSf}. In sum, similar to the single-component cases, the current AMS-02 data on the positron fraction seems to be best fitted with a relatively small heavy-DM cutoff, which is in mild tension with the excesses at higher energies in the total $e^+ + e^-$ flux. But all the benchmarks can give good enough fit to the AMS-02 data, which cannot be achieved by the single-DM models with the cutoff larger than 1 TeV.
{\begin{table}[h]
\caption{Parameters leading to the minimal values of $\chi^2$ with the cutoffs of heavy DM being 600, 800, 1200, and 1500 GeV, respectively.}
{\begin{tabular}{@{}ccccccccc@{}} \toprule
$E_{cH}({\rm GeV})$& $\kappa_1$ & $\kappa_2$ & $\epsilon^{e}_{H,L}$& $\epsilon^{\mu}_{H,L}$  & $\epsilon^{\tau}_{H,L}$
 & $\tau_{H,L}(10^{26}{\rm s})$ & $\chi^2_{\rm min}$ & $\chi^2_{\rm min}/{\rm d.o.f.}$
\\
\colrule
600 & 0.94 & 1.49 & 0.18, 0.02  & 0.74, 0.00 &  0.08, 0.98 & 1.06, 0.93 & 102 & 0.78 \\
800 & 0.94 & 1.49 & 0.04, 0.02  & 0.65, 0.00 &  0.31, 0.98 & 0.75, 0.97 & 102 & 0.78 \\
1200 & 0.94 & 1.50 & 0.00, 0.01  & 0.80, 0.00 &  0.20, 0.99 & 0.43, 1.12 & 102 & 0.78\\
1500 & 0.94 & 1.50 & 0.00, 0.04  & 1.00, 0.17 &  0.00, 0.79 & 0.42, 1.39 & 102 & 0.78\\
\botrule
\end{tabular}\label{tab_2DM}}
\end{table}
}

{Now we hope to make clear the role of the DM masses $M_i$ and the electron/positron cutoffs $E_{ci}$ played in our fit. In the present paper we consider the decay process $\chi_i \to \ell^\pm Y^\mp$ for each of the DM components $\chi_i$ with a unique $Y^\mp$. The existence of extra particle $Y^\pm$ breaks the degeneracy between the electron cutoff $E_{ci}$ and the DM mass $M_i=2E_{ci}$ in the conventional modes $\chi_i\to \ell^+ \ell^-$. Instead, they obey the new relation specified in Eq.~(\ref{cutoff}). In other words, they are totally independent when $M_Y$ is free. Note that $E_{ci}$ and $M_i$ have different effects on the predicted injection spectra. Specifically, the DM cutoffs $E_{ci}$ affect the shape of the final spectra by determining the energy scale where the injected $e^\pm$ fluxes drop at, while the DM masses $M_i$ enters the spectra only through the DM $e^\pm$ source terms in the product with the DM lifetimes $\tau_i$ in Eq.~(\ref{source}). To put it in another way, the product $\tau_i M_i$ is the only independent parameter that the fitting procedure can determine. In this sense, the DM lifetimes get their values and meanings by specifying the DM masses. Also note that the goodness of the fit is essentially controlled by the overall normalization factor $\tau_i M_i$ and the shape of the spectra which is in turn closely related to the cutoffs $E_{ci}$ and the decay modes considered. Therefore, the variation of the DM masses $M_i$ alone will not change the goodness of the fit, {\it i.e.}, the value of the minimum $\chi^2$. Rather, we only need to tune the DM lifetimes $M_i$ to make the combination $M_i \tau_i$ constant.}

\section{Remarks on the Diffuse $\gamma$-ray Constraints}
\label{s4}

Finally, we would make some comments on the diffuse $\gamma$-ray constraints in the present 
single- and two-component decaying DM scenarios. As pointed in 
Refs.~\cite{Cirelli:2012ut, Papucci:2009gd, Beacom:2004pe, Gamma_DM, Gamma_extraDMIC, Gamma_extra, Ibarra:2007wg,Ishiwata:2009vx}, 
the current diffuse $\gamma$-ray measurement by Fermi-LAT has already excluded a large range of the parameter space of the single-component leptophilic decaying DM models trying to explain the positron/electron excesses. However, it has been shown in Refs.~\cite{2DM_1,2DM_2} that the present two-component decaying DM scenario is promising to reconcile the tension between these two kinds of experiments, in which the prediction of the diffuse $\gamma$-ray spectrum is done by summing all the contributions to the background and DM signals. In the following, we shall argue that this feature persists for the results in Tables~\ref{tab_1DM} and~\ref{tab_2DM}. Since the final predictions of the diffuse $\gamma$-ray spectrum are similar to those shown in Refs.~\cite{2DM_1,2DM_2}, we shall not repeat such calculation again. Instead, we would like to reach this conclusion by arguing the reasons behind.

Refs.~\cite{Cirelli:2012ut,Papucci:2009gd} have made the detailed discussions of the diffuse $\gamma$-ray constraints on
 the single-component decaying DM models with the conventional decay channels, representing the standard references in the literature. 
 Our present study is mainly based on the comparison between our scenario with these two papers. 
First of all, the interpretation of the Fermi-LAT diffuse $\gamma$-ray data in Ref.~\cite{Cirelli:2012ut},
which assumes that the measured spectrum should be fitted with a simple power law function arising from the conventional astrophysical sources,
 is very different from our viewpoint. The possible contribution from DM could only be manifested as the residue after the subtraction of the data to this background, leading to very stringent DM lifetime bounds. From our perspective, however, the measured spectrum is the total summation of the astrophysical background and the signals from DM decays. Therefore, the constraints in Ref.~\cite{Cirelli:2012ut} cannot be applied to our cases.

On the other hand, the constraints from Ref.~\cite{Papucci:2009gd} are more relevant to our present scenario since the authors, Papucci and Strumia (PS), did not assume any astrophysical background in their derivation. The bounds $\tau^{PS}$ for various decay channels are shown in Fig.~8 in Ref.~\cite{Papucci:2009gd}, from which we can read off the lowest DM lifetime bounds for the corresponding DM masses. However, these DM lifetime bounds have to be transformed before they can be used here. One prominent difference lies in that in our scenario we have $N$ components with an equal amount DM density by assumption, so that there is a factor $1/N$ suppression for each channel. Moreover, the DM decay processes  in this paper have only one lepton in the final states, rather than a lepton pair in the usual models in Ref.~\cite{Papucci:2009gd}, so that additional 1/2 suppression should be also taken into account. Another aspect is that the DM masses in our scenario are different from those in the lepton pair decay processes in which $m_{\rm DM}^{PS} = 2 E_c$. By considering all these effects, we can transform the DM lifetime bounds shown in Ref.~\cite{Papucci:2009gd} into those for our models via
\begin{eqnarray}\label{DMbound}
\tau_l=\frac{M_{\rm DM}^{PS}\tau_l^{PS}}{2 N M_i},
\end{eqnarray}
where the subscript $l$ denotes the corresponding lepton channel.

For the single-component DM models  in Table~\ref{tab_1DM}, the dominant decay channels are all $\tau$ modes. Since the DM cutoffs are 600, 800, 1000, and 1500 GeV, the corresponding lifetime bounds for the tau-pair final state
lie in the range $2\sim 3\times 10^{26}$~s, from which the lifetime bounds for our scenario can be obtained via Eq.~(\ref{DMbound}) as $1\sim 1.5 \times 10^{26}$~s. Obviously, the best-fit lifetimes in Table~\ref{tab_1DM} are already excluded by these bounds. Therefore, it is seen that the single-component DM models used to explain the AMS-02 excesses have some kind of tension with the Fermi-LAT diffuse $\gamma$-ray results.

However, our two-component DM models do not possess this problem. For example, the light DM with $E_{cL}=100$~GeV predominantly decays 
via the $\tau$-channel as shown in Table~\ref{tab_2DM}. The relevant lifetime bound in Ref.~\cite{Papucci:2009gd} is $\tau^{PS}_\tau = 1.5\times10^{26}$~s for ${\rm DM} \to \tau^+\tau^-$ with  $M^{PS}_{\rm DM}=200$~GeV, which corresponds to $\tau_\tau = 2\times 10^{25}$~s 
with the light DM mass $M_L=416$~GeV. The same argument can also lead us to the heavy DM lifetime bounds $\tau_\mu = 0.75\sim 1.25\times 10^{25}$~s for the dominant $\mu$ channels with $E_{cH} = 600 \sim 1500$~GeV. It is clear that these 
 bounds are still much lower than the two best-fit DM lifetimes in all of the four benchmarks listed in Table~\ref{tab_2DM}, from which we can obtain the conclusion that the two-component decaying DM models are more favorable than their single-component cousins by the Fermi-LAT diffuse $\gamma$-ray data.

\section{Conclusion}
\label{s5}
The recent release of the AMS-02 data on the positron fraction and electron/positron respective fluxes has given 
us some new hints toward the DM interpretation of the positron/electron excesses. 
In the present paper, we have revisited the multi-component decaying DM scenario introduced in our previous work~\cite{2DM_1,2DM_2} with the updated AMS-02 datasets. It is found that both  single- and two-component DM models can yield consistent fits to the aforementioned datasets, with the two-component cases even better. The hardening behavior in $e^+/e^-$ fluxes around 30 GeV can be explained by the transition from the background-dominated to the DM-signal regions. For the single-component DM models, the AMS-02 data, especially the positron fraction, constrain the dominant DM decay channel to be the $\tau$-mode with its cutoff  lighter than 1 TeV,
resulting in  that the total $e^+ + e^-$ flux stops excessing too early to explain the data. 
In comparison, the two-component DM models provide even better fit to the AMS-02 data, 
in which the heavy DM decays predominately via the $\mu-$channel, while the light one with $E_{cL}=100$~GeV mostly via the $\tau$-channel. 
We have also made some comments on the diffuse $\gamma$-ray constraint from the Fermi-LAT measurement. We have found that the corresponding dataset has already excluded the best-fit lifetimes of the single-component DM models with the dominant $\tau$ decay channels, while still allows the two-component DM models benchmarks listed in Table~\ref{tab_2DM}. In sum, the two-component DM models are more favored by the current indirect DM searches, 
 providing a better fit to the AMS-02 $e^+/e^-$ data, which are also in good agreement with the Fermi-LAT diffuse $\gamma$-ray data.

Moreover, our best-fit parameters with the heavy DM's cutoff $E_{cH} = 800$~GeV predicts the decline tendency
above ~300 GeV claimed in Ref.~\cite{AMSf}, while a heavy DM with $E_{cH} = 1200$ or 1500 GeV can give a better description 
of the high energy behavior of the AMS-02 $e^+ + e^-$ flux data. However, there is no model to accommodate both high-energy features, 
 regarded as some tensions among the AMS-02 datasets. We hope that the more precise AMS-02 data in the near future can settle 
down this problem.

\section*{Acknowledgments}
The work was supported in part by National Center for Theoretical Science, National Science
Council (NSC-101-2112-M-007-006-MY3) and National Tsing Hua
University (103N2724E1).



\begin{thebibliography}{0}

\bibitem{AMSf}
  L.~Accardo {\it et al.}  [AMS Collaboration],
  Phys.\ Rev.\ Lett.\  {\bf 113}, 121101 (2014).

\bibitem{AMSep}
  M.~Aguilar {\it et al.}  [AMS Collaboration],
  Phys.\ Rev.\ Lett.\  {\bf 113}, 121102 (2014).





\bibitem{AMS01}
  M.~Aguilar {\it et al.}  [AMS-01 Collaboration],
  Phys.\ Lett.\ B {\bf 646}, 145 (2007);
\bibitem{AMS02}
  M.~Aguilar {\it et al.}  [AMS Collaboration],
  Phys.\ Rev.\ Lett.\  {\bf 110}, 141102 (2013).

\bibitem{ATIC}
  J.~Chang {\it et al.},
  Nature {\bf 456}, 362 (2008).

\bibitem{PAMELA}
  O.~Adriani {\it et al.}  [PAMELA Collaboration],
  Nature {\bf 458}, 607 (2009).


\bibitem{PAMELA2}
  O.~Adriani {\it et al.}  [ PAMELA Collaboration],
  arXiv:1308.0133 [astro-ph.HE].

\bibitem{FermiLAT}
  A.~A.~Abdo {\it et al.}  [Fermi LAT Collaboration],
  Phys.\ Rev.\ Lett.\  {\bf 102}, 181101 (2009).

\bibitem{FermiLAT1}
  M.~Ackermann {\it et al.}  [Fermi LAT Collaboration],
  Phys.\ Rev.\ D {\bf 82}, 092004 (2010).

\bibitem{FermiLATp}
  M.~Ackermann {\it et al.}  [Fermi LAT Collaboration],
  Phys.\ Rev.\ Lett.\  {\bf 108}, 011103 (2012).


\bibitem{pulsar}
  S.~Profumo,
  Central Eur.\ J.\ Phys.\  {\bf 10}, 1 (2011);
  T.~Linden and S.~Profumo,
  Astrophys.\ J.\  {\bf 772}, 18 (2013);
  P.~F.~Yin {\it et al.},
  Phys.\ Rev.\ D {\bf 88}, 023001 (2013);
  D.~Gaggero {\it et al.},
  Phys.\  Rev.\  Lett.\  111, {\bf 021102} (2013);
  C.~Venter, A.~Kopp, P.~L.~Gonthier, A.~K.~Harding and I.~B¨¹sching,
  arXiv:1410.6462 [astro-ph.HE].

\bibitem{Lin:2014vja} 
  S.~J.~Lin, Q.~Yuan and X.~J.~Bi,
  Phys.\ Rev.\ D {\bf 91}, no. 6, 063508 (2015)
  [arXiv:1409.6248 [astro-ph.HE]].

\bibitem{Boudaud:2014dta}
  M.~Boudaud, S.~Aupetit, S.~Caroff, A.~Putze, G.~Belanger, Y.~Genolini, C.~Goy and V.~Poireau {\it et al.},
  arXiv:1410.3799 [astro-ph.HE].


\bibitem{DMindependent}
  K.~Ishiwata, S.~Matsumoto and T.~Moroi,
  Phys.\ Lett.\ B {\bf 675}, 446 (2009);
  L.~Bergstrom {\it et al.},
  Phys.\ Rev.\ Lett.\  {\bf 111}, 171101 (2013);
  D.~Gaggero and L.~Maccione,
  JCAP {\bf 1312}, 011 (2013)

\bibitem{annihilation}
  L.~Bergstrom, T.~Bringmann and J.~Edsjo,
  Phys.\ Rev.\ D {\bf 78}, 103520 (2008);
  M.~Cirelli and A.~Strumia,
  PoS IDM {\bf 2008}, 089 (2008);
  E.~Nezri, M.~H.~Tytgat and G.~Vertongen,
  JCAP {\bf 0904}, 014 (2009);
  X.~J.~Bi {\it et al.},
  JHEP {\bf 0904}, 103 (2009);



\bibitem{annihilationAMS}
  P.~S.~Dev {\it et al.},
  arXiv:1307.6204 [hep-ph];
  L.~Feng {\it et al.},
  Phys.\ Lett.\ B {\bf 728}, 250 (2014);
  Q.~H.~Cao, C.~R.~Chen and T.~Gong,
  arXiv:1409.7317 [hep-ph].
  
\bibitem{AnnihilationDecay}
  K.~Cheung, P.~Y.~Tseng and T.~C.~Yuan,
  Phys.\ Lett.\ B {\bf 678}, 293 (2009).



\bibitem{Jin:2014ica}
  H.~B.~Jin, Y.~L.~Wu and Y.~F.~Zhou,
  arXiv:1410.0171 [hep-ph].





\bibitem{decay}
  C.~R.~Chen and F.~Takahashi,
  JCAP {\bf 0902}, 004 (2009);
  P.~F.~Yin {\it et al.},
  Phys.\ Rev.\ D {\bf 79}, 023512 (2009);
  K.~Hamaguchi {\it et al.},
  Phys.\ Lett.\ B {\bf 674}, 299 (2009);
  A.~Ibarra and D.~Tran,
  JCAP {\bf 0902}, 021 (2009);
  E.~Nardi, F.~Sannino and A.~Strumia,
  JCAP {\bf 0901}, 043 (2009);
  I.~Gogoladze {\it et al.},
  Phys.\ Rev.\ D {\bf 79}, 055019 (2009);
  S.~L.~Chen {\it et al.},
  Phys.\ Lett.\ B {\bf 677}, 311 (2009);
  A.~Arvanitaki {\it et al.},
  Phys.\ Rev.\ D {\bf 80}, 055011 (2009);
  H.~Fukuoka, J.~Kubo and D.~Suematsu,
  Phys.\ Lett.\ B {\bf 678}, 401 (2009);

\bibitem{Ishiwata:2009vx}
  K.~Ishiwata, S.~Matsumoto and T.~Moroi,
  JHEP {\bf 0905}, 110 (2009).

\bibitem{decayAMS}
  A.~Ibarra, D.~Tran and C.~Weniger,
  Int.\ J.\ Mod.\ Phys.\ A {\bf 28}, 1330040 (2013);
  A.~Ibarra, A.~S.~Lamperstorfer and J.~Silk,
  arXiv:1309.2570 [hep-ph];
  M.~Ibe, S.~Matsumoto, S.~Shirai and T.~T.~Yanagida,
  arXiv:1409.6920 [hep-ph].

\bibitem{Ko:2014lsa}
  P.~Ko and Y.~Tang,
  arXiv:1410.7657 [hep-ph].

\bibitem{3bodydecay}
  A.~Arvanitaki {\it et al.},
  Phys.\ Rev.\ D {\bf 79}, 105022 (2009);
  K.~Hamaguchi, S.~Shirai and T.~T.~Yanagida,
  Phys.\ Lett.\ B {\bf 673}, 247 (2009);
  C.~H.~Chen, C.~Q.~Geng and D.~V.~Zhuridov,
  Phys.\ Lett.\ B {\bf 675}, 77 (2009).

\bibitem{3bodydecayAMS}
  M.~Ibe {\it et al.},
  JHEP {\bf 1307}, 063 (2013);
  K.~Kohri and N.~Sahu,
  Phys.\ Rev.\ D {\bf 88}, 103001 (2013).

\bibitem{Chen:2009gd}
  C.~H.~Chen, C.~Q.~Geng and D.~V.~Zhuridov,
  JCAP {\bf 0910}, 001 (2009).

\bibitem{2body}
  V.~Barger {\it et al.},
  Phys.\ Lett.\ B {\bf 672}, 141 (2009);
  M.~Cirelli {\it et al.},
  Nucl.\ Phys.\ B {\bf 813}, 1 (2009);
  C.~R.~Chen, F.~Takahashi and T.~T.~Yanagida,
  Phys.\ Lett.\ B {\bf 673}, 255 (2009);
  C.~R.~Chen {\it et al.},
  Prog.\ Theor.\ Phys.\  {\bf 122}, 553 (2009);
  J.~Liu, P.~F.~Yin and S.~H.~Zhu,
  Phys.\ Rev.\ D {\bf 79}, 063522 (2009).

\bibitem{2bodyAMSa}
  A.~Sharma,
  arXiv:1304.0831 [astro-ph.CO];
  J.~Kopp,
  Phys.\ Rev.\ D {\bf 88}, 076013 (2013);
  A.~De Simone, A.~Riotto and W.~Xue,
  JCAP {\bf 1305}, 003 (2013);
  I.~Cholis and D.~Hooper,
  Phys.\ Rev.\ D {\bf 88}, 023013 (2013);
  L.~Feng and Z.~Kang,
  JCAP {\bf 1310}, 008 (2013);
  Q.~Yuan and X.~J.~Bi,
  Phys.\ Lett.\ B {\bf 727}, 1 (2013).
  Y.~Kajiyama, H.~Okada and T.~Toma,
  Eur.\ Phys.\ J.\ C {\bf 74}, 2722 (2014)
  [arXiv:1304.2680 [hep-ph]];
  Q.~Yuan, X.~J.~Bi, G.~M.~Chen, Y.~Q.~Guo, S.~J.~Lin and X.~Zhang,
  Astropart.\ Phys.\  {\bf 60}, 1 (2014)
  [arXiv:1304.1482 [astro-ph.HE]];
  H.~B.~Jin, Y.~L.~Wu and Y.~F.~Zhou,
  JCAP {\bf 1311}, 026 (2013).

\bibitem{PAMELApr}
  O.~Adriani {\it et al.}  [PAMELA Collaboration],
  Phys.\ Rev.\ Lett.\  {\bf 105}, 121101 (2010).
  
\bibitem{AMSpr} AMS-02 Collaboration, Talks at the `AMS Days at CERN', 15-17 April 2015.

\bibitem{Giesen:2015ufa} 
  G.~Giesen, M.~Boudaud, Y.~Genolini, V.~Poulin, M.~Cirelli, P.~Salati, P.~D.~Serpico and J.~Feng {\it et al.},
  arXiv:1504.04276 [astro-ph.HE].

\bibitem{Jin:2015sqa} 
  H.~B.~Jin, Y.~L.~Wu and Y.~F.~Zhou,
  arXiv:1504.04604 [hep-ph].


{
\bibitem{astrophysical} 
  K.~Blum, B.~Katz and E.~Waxman,
  Phys.\ Rev.\ Lett.\  {\bf 111}, no. 21, 211101 (2013)
  [arXiv:1305.1324 [astro-ph.HE]];
  S.~P.~Ahlen and G.~Tarlé,
  arXiv:1410.7239 [astro-ph.HE].
}

\bibitem{Dado:2015lta} 
  S.~Dado and A.~Dar,
  arXiv:1504.03261 [astro-ph.HE].

\bibitem{2DM_1}
  C.~Q.~Geng, D.~Huang and L.~H.~Tsai,
  Phys.\ Rev.\ D {\bf 89},  055021 (2014)
  [arXiv:1312.0366 [hep-ph]].

\bibitem{2DM_2}
  C.~Q.~Geng, D.~Huang and L.~H.~Tsai,
  Mod. Phys. Lett. A {\bf 29}, 1440003 (2014)
  [arXiv:1405.7759 [hep-ph]].

\bibitem{2DM_others} Other aspects of multi-component dark matter models are studied in {\emph e.g.}, 
  K.~R.~Dienes and B.~Thomas,
  Phys.\ Rev.\ D {\bf 85}, 083523 (2012)
  [arXiv:1106.4546 [hep-ph]];
  Phys.\ Rev.\ D {\bf 85}, 083524 (2012)
  [arXiv:1107.0721 [hep-ph]];
  K.~R.~Dienes, J.~Kumar and B.~Thomas,
  Phys.\ Rev.\ D {\bf 88}, no. 10, 103509 (2013)
  [arXiv:1306.2959 [hep-ph]];
  P.~H.~Gu,
  Phys.\ Dark Univ.\  {\bf 2}, 35 (2013)
  [arXiv:1301.4368 [hep-ph]];
  Y.~.B.~Zeldovich {\it et al.},
  Sov.\ J.\ Nucl.\ Phys.\  {\bf 31}, 664 (1980);
R.~V.~Konoplich and M.~Yu.~Khlopov, Phys.\ Atom.\ Nucl.\  {\bf 57}, 425 (1994);
D.~Fargion {\it et al.}, Phys.\ Rev.\ {\bf D52}, 1828 (1995);
ASTRODAMUS collaboration, In: Proc.~1 Int.~Conf.~on cosmoparticle physics "Cosmion-94", dedic. to 80 Anniv. of Ya.B. Zeldovich and 5 Mem. of A.D.Sakharov, Moscow, Dec. 5-14, 1994. Eds. M.Yu. Khlopov {\it et al.}, Editions Frontieres, 1996. PP. 99-106 and 107-112;
  K.~Belotsky {\it et al.},
  Phys.\ Atom.\ Nucl.\  {\bf 71}, 147 (2008);
  K.~Belotsky, M.~Khlopov, C.~Kouvaris and M.~Laletin,
  Adv.\ High Energy Phys.\  {\bf 2014}, 214258 (2014)
  [arXiv:1403.1212 [astro-ph.CO]];
  K.~Belotsky, M.~Khlopov and M.~Laletin,
  arXiv:1411.3657 [hep-ph];
{K.~M.~Zurek,
  Phys.\ Rev.\ D {\bf 79}, 115002 (2009)
  [arXiv:0811.4429 [hep-ph]];
  M.~Aoki, M.~Duerr, J.~Kubo and H.~Takano,
  Phys.\ Rev.\ D {\bf 86}, 076015 (2012)
  [arXiv:1207.3318 [hep-ph]];
  D.~Chialva, P.~S.~B.~Dev and A.~Mazumdar,
  Phys.\ Rev.\ D {\bf 87}, no. 6, 063522 (2013)
  [arXiv:1211.0250 [hep-ph]];
  S.~Bhattacharya, A.~Drozd, B.~Grzadkowski and J.~Wudka,
  JHEP {\bf 1310}, 158 (2013)
  [arXiv:1309.2986 [hep-ph]];
  K.~R.~Dienes, J.~Kumar, B.~Thomas and D.~Yaylali,
  Phys.\ Rev.\ Lett.\  {\bf 114}, no. 5, 051301 (2015)
  [arXiv:1406.4868 [hep-ph]];
}

\bibitem{FermiLAT_Gamma}
  M.~Ackermann {\it et al.}  [LAT Collaboration],
  Phys.\ Rev.\ D {\bf 86}, 022002 (2012).

\bibitem{HESS} 
  F.~Aharonian {\it et al.}  [HESS Collaboration],
  Phys.\ Rev.\ Lett.\  {\bf 101}, 261104 (2008)
  [arXiv:0811.3894 [astro-ph]]; 
  Astron.\ Astrophys.\  {\bf 508}, 561 (2009)
  [arXiv:0905.0105 [astro-ph.HE]].



\bibitem{SNR}
  R.~Trotta {\it et al.},
  Astrophys.\ J.\  {\bf 729}, 106 (2011).

\bibitem{GALPROP}
  A.~W.~Strong and I.~V.~Moskalenko,
  Astrophys.\ J.\  {\bf 509}, 212 (1998).

\bibitem{CMBpower}
  X.~L.~Chen and M.~Kamionkowski,
  Phys.\ Rev.\ D {\bf 70}, 043502 (2004)
  [astro-ph/0310473];
  T.~R.~Slatyer,
  Phys.\ Rev.\ D {\bf 87}, no. 12, 123513 (2013)
  [arXiv:1211.0283 [astro-ph.CO]];
  J.~M.~Cline and P.~Scott,
  JCAP {\bf 1303}, 044 (2013)
  [Erratum-ibid.\  {\bf 1305}, E01 (2013)]
  [arXiv:1301.5908 [astro-ph.CO]].

\bibitem{NFW}
  J.~F.~Navarro, C.~S.~Frenk and S.~D.~M.~White,
  Astrophys.\ J.\  {\bf 490}, 493 (1997)
  [astro-ph/9611107];
  A.~F.~Neto, L.~Gao, P.~Bett, S.~Cole, J.~F.~Navarro, C.~S.~Frenk, S.~D.~M.~White and V.~Springel {\it et al.},
  Mon.\ Not.\ Roy.\ Astron.\ Soc.\  {\bf 381}, 1450 (2007)
  [arXiv:0706.2919 [astro-ph]].

\bibitem{PYTHIA}
  T.~Sjostrand, S.~Mrenna and P.~Z.~Skands,
  JHEP {\bf 0605}, 026 (2006).

\bibitem{DiffuseEquation}
  E.~A.~Baltz and J.~Edsjo,
  Phys.\ Rev.\ D {\bf 59}, 023511 (1998).

\bibitem{SolarModulation}
  L.~J.~Gleeson and W.~I.~Axford,
  Astrophys.\ J.\  {\bf 154}, 1011 (1968).
  
\bibitem{isothermal}
  K.~G.~Begeman, A.~H.~Broeils and R.~H.~Sanders,
  Mon.\ Not.\ Roy.\ Astron.\ Soc.\  {\bf 249}, 523 (1991).

\bibitem{AMSt} Y.~H.~Chang, talk given at 2nd International Workshop on Particle Physics and Cosmology after Higgs and Planck,\\ http://www.phys.nthu.edu.tw/~dark/higplk2014/doc/9/sec\_i/AMS\_PPCHP.pdf

\bibitem{Cirelli:2012ut}
  M.~Cirelli {\it et al.}, E.~Moulin, P.~Panci, P.~D.~Serpico and A.~Viana,
  Phys.\ Rev.\ D {\bf 86}, 083506 (2012).

\bibitem{Papucci:2009gd}
 M.~Papucci and A.~Strumia,
  JCAP {\bf 1003}, 014 (2010).


\bibitem{Beacom:2004pe}
  J.~F.~Beacom, N.~F.~Bell and G.~Bertone,
  Phys.\ Rev.\ Lett.\  {\bf 94}, 171301 (2005).


\bibitem{Gamma_DM}
  R.~Essig, N.~Sehgal and L.~E.~Strigari,
  Phys.\ Rev.\ D {\bf 80}, 023506 (2009).

\bibitem{Gamma_extraDMIC}
  A.~A.~Abdo {\it et al.}  [Fermi-LAT Collaboration],
  Phys.\ Rev.\ Lett.\  {\bf 104}, 101101 (2010);
  M.~Cirelli and P.~Panci,
  Nucl.\ Phys.\ B {\bf 821}, 399 (2009);
  S.~Matsumoto, K.~Ishiwata and T.~Moroi,
  Phys.\ Lett.\ B {\bf 679}, 1 (2009).

\bibitem{Gamma_extra}
  A.~Ibarra, D.~Tran and C.~Weniger,
  JCAP {\bf 1001}, 009 (2010).
  C.~R.~Chen, F.~Takahashi and T.~T.~Yanagida,
  Phys.\ Lett.\ B {\bf 671}, 71 (2009).

\bibitem{Ibarra:2007wg}
  A.~Ibarra and D.~Tran,
  Phys.\ Rev.\ Lett.\  {\bf 100}, 061301 (2008).


\end{thebibliography}
\end{document}